\documentclass{ws-procs961x669}            
\begin{document}
\title{Closing the cosmological loop with the redshift drift}

\author{C. J. A. P. Martins$^*$}
\address{Centro de Astrof\'{\i}sica da Universidade do Porto, and\\
Instituto de Astrof\'{\i}sica e Ci\^encias do Espa\c co, Universidade do Porto,\\
Rua das Estrelas, 4150-762 Porto, Portugal\\
$^*$E-mail: Carlos.Martins@astro.up.pt}

\author{C. S. Alves}
\address{Department of Physics and Astronomy, University College London,\\
Gower Street, London WC1E 6BT, United Kingdom, and\\
Centro de Astrof\'{\i}sica da Universidade do Porto,\\
Rua das Estrelas, 4150-762 Porto, Portugal}

\author{J. Esteves}
\address{Université de Montpellier,\\ 163 rue Auguste Broussonnet, 34090 Montpellier, and\\
Centro de Astrof\'{\i}sica da Universidade do Porto,\\
Rua das Estrelas, 4150-762 Porto, Portugal}

\author{A. Lapel}
\address{Institut d'Astrophysique de Paris,\\
98bis Bd Arago, 75014 Paris, France and\\
Centro de Astrof\'{\i}sica da Universidade do Porto,\\
Rua das Estrelas, 4150-762 Porto, Portugal}

\author{B. G. Pereira}
\address{Faculdade de Ci\^encias, Universidade do Porto,\\
Rua do Campo Alegre, 4150-007 Porto, Portugal, and\\
Centro de Astrof\'{\i}sica da Universidade do Porto,\\
Rua das Estrelas, 4150-762 Porto, Portugal}

\begin{abstract}
The redshift drift (also known as the Sandage Test) is a model-independent probe of fundamental cosmology, enabling us to watch the universe expand in real time, and thereby to confirm (or not) the recent acceleration of the universe without any model-dependent assumptions. On the other hand, by choosing a fiducial model one can also use it to constrain the model parameters, thereby providing a consistency test for results obtained with other probes. The drift can be measured by the Extremely Large Telescope and also by the full SKA. Recently two alternative measurement methods have been proposed: the cosmic accelerometer, and the differential redshift drift. Here we summarize a comparative analysis of the various methods and their possible outcomes, using both Fisher Matrix and MCMC techniques. We find that no single method is uniformly better than the others. Instead, their comparative performance depends both on experimental parameters (including the experiment time and redshift at which the measurement is made) and also on the scientific goal (e.g., detecting the drift signal with high statistical significance, constraining the matter density, or constraining the dark energy properties). In other words, the experiment should be optimized for the preferred scientific goal.
\end{abstract}


\bodymatter

\section{Introduction}

The observational evidence for the acceleration of the universe shows that our canonical theories of cosmology and particle physics are at least incomplete, and possibly incorrect. New physics is out there, waiting to be discovered; we must search for, identify and characterize this new physics. But so far, the interpretation of all our extant cosmological observations is model-dependent, in the sense that it requires a number of assumptions on an underlying cosmological model. Ideally, one would like some fully model-independent observables, enabling consistency tests of these underling assumptions. And indeed there is one conceptually simple, but operationally very challenging, such observable: the redshift drift of objects following the cosmological expansion.

The CosmoESPRESSO team uses the universe as a laboratory to address, with precision spectroscopy and other observational, computational and theoretical tools, grand-challenge questions including the universality of physical laws, the characterization of the large-scale behaviour of gravity (and specifically of the source of the recent acceleration of the universe) and the study of fossil relics or early stages of the universe's evolution, together with contributions to ESO and ESA next-generation facilities and to education and public understanding of science. In what follows we highlight recent contributions of the CosmoESPRESSO team to this fundamental quest, pertaining to the cosmological impact of forthcoming redshift drift measurements. Further details can be found in Ref. \citenum{drift1} and Ref. \citenum{drift2}.

\section{The redshift drift and its sensitivity}

The idea that the redshift of objects following the cosmological expansion changes with time---known as the redshift drift--- is many decades old \cite{Sandage,Mcvittie}. Conceptually, a measurement of the redshift drift is fundamentally different from our other astrophysical observables. In our observations at cosmological distances done so far, we are mapping our present-day past light cone. Instead, the redshift drift allows us to compare different past light cones. More crudely, this corresponds to watching the Universe expand in real time. Other than being a different probe of the universe, its conceptual importance is that it is a model-independent probe of the expansion of the universe, making no assumption on geometry, clustering or the behaviour of gravity, and therefore of crucial importance for fundamental cosmology.

The practical difficulty is that cosmologically relevant timescales are orders of magnitude larger than human ones: a measurement of the redshift drift therefore requires exquisite sensitivity, several orders of magnitude better than currently available: the best currently available bound \cite{Darling}, is about three orders of magnitude larger than the signal expected for the standard $\Lambda$CDM cosmology with reasonable choices of its model parameters, and contains systematic errors comparable to or larger than the statistical ones. Nevertheless, measuring this drift is a key science and design driver for the development of  the Extremely Large Telescope (ELT), and specifically for its high-resolution spectrograph currently known as ELT-HIRES \citep{HIRES}.

A detailed feasibility study of high-redshift measurements by the ELT, on a decade timescale, was done in Ref. \citenum{Liske}, who also discussed possible targets. It is important to note that although ELT redshift drift measurements, on their own, lead to cosmological parameter constraints that are not tighter than those available by more classical probes (such as supernovae or the CMB) they do probe regions of parameter space that are different from (and sometimes actually orthogonal to) those of other probes, enabling the breaking of degeneracies and therefore leading to more stringent combined constraints \cite{drift1}; we will provide an explicit demonstration later in this contribution. More recently, it has also been pointed out that measurements at low redshifts can in principle be made by the full Square Kilometre Array (SKA) \cite{Klockner}, while measurements at intermediate redshifts can be done by intensity mapping measurements such as CHIME \cite{Chime} and HIRAX \cite{HIRAX}. Measurements of all three of these will require appropriate hardware configurations.

Redshift drift measurements are a key part of what is commonly called real-time cosmology \cite{Quercellini}. Usually they rely on the first derivative of the redshift. Recently the role of second derivatives of the redshift, which should also be within the reach of the full SKA, has been studied by Ref. \cite{Second}. Fisher Matrix techniques \cite{FMA1,FMA2} are well suited for a comparative study of the cosmological impact of redshift drift measurements by these facilities; such a study is described in detail in Ref. \citenum{drift1}.

The redshift drift of an astrophysical object following the cosmological expansion, for an observer looking at it over a time span $\Delta t$, is
\begin{equation}
\frac{\Delta z}{\Delta t}=H_0 \left[1+z-E(z)\right]\,,
\end{equation}
where for convenience we have defined the dimensionless Hubble parameter $E(z)=H(z)/H_0$, although the actual astrophysical observable is usually a spectroscopically measured velocity
\begin{equation}
\Delta v=\frac{c\Delta z}{1+z}=(cH_0\Delta t)\left[1-\frac{E(z)}{1+z}\right]\,.
\end{equation}
The dependence on the Hubble parameter naturally leads to a model-dependent redshift dependence of the drift. Broadly speaking, in a universe that is currently accelerating but was deccelerating in the past the drift will be positive at low redshifts and negative for higher redshifts, while in a universe that always deccelerates the redshift drift would always be negative. To give a specific example, for a flat $\Lambda$CDM model, apart from trivially vanishing at $z=0$, the signal will also vanish at
\begin{equation}
z_{zero}=\frac{1-3\Omega_m+\sqrt{1+2\Omega_m-3\Omega^2_m}}{2\Omega_m}\,,
\end{equation}
while the redshift of maximum spectroscopic velocity is
\begin{equation}
z_{v,max}=\left[\frac{2(1-\Omega_m)}{\Omega_m}\right]^{1/3}-1
\end{equation}
and the maximum of the drift itself is obtained by solving the quartic equation
\begin{equation}
9\Omega_m^2(1+z)^4=4\Omega_m(1+z)^3+4(1-\Omega_m)\,.
\end{equation}
More specifically, if we choose $\Omega_m=0.3$ (in agreement with contemporary cosmological data) we obtain $z_{zero}=\sim 2.09$, $z_{v,max}\sim 0.67$ and $z_{z,max}\sim 0.95$. It is interesting to note that for flat $\Lambda$CDM with $\Omega_m=0.3$, the acceleration phase does start at redshift $z_{v,max}$, while the cosmological constant only starts dominating the Friedmann equation at
\begin{equation}
z_{\Lambda}=\left[\frac{(1-\Omega_m)}{\Omega_m}\right]^{1/3}-1\,,
\end{equation}
which for $\Omega_m$ leads to $z_{\Lambda}\sim0.33$.

We can analytically explore the sensitivity of the redshift drift to the cosmological parameters, for a fiducial flat CPL model \cite{CPL1,CPL2}. In this case we can write the Friedmann equation as
\begin{equation}
E^2(z)=\Omega_m(1+z)^3+\Omega_\phi(1+z)^{3(1+w_0+w_a)}\exp{\left[-\frac{3w_az}{1+z}\right]}\,.
\end{equation}
It is also convenient to define a dimensionless redshift drift
\begin{equation}
S_z=\frac{1}{H_{100}}\frac{\Delta z}{\Delta t}=h\left[1+z-E(z)\right]\,,
\end{equation}
where we have defined $H_0=hH_{100}$ with $H_{100}=100$ km/s/Mpc; the corresponding observable spectroscopic velocity will be denoted
\begin{equation}
S_v={\Delta v}=kh\left[1-\frac{E(z)}{1+z}\right]\,.
\end{equation}
In these we have introduced $k=cH_{100}\Delta t$, which is a constant parameter for a given observation time, with units of cm/s: for $\Delta t=1$ year, $k=3.064$ cm/s. This provides a rough estimate of the magnitude of the signal in the relevant time span, and therefore also an estimate of the required sensitivity of the spectroscopic measurements. Then for each cosmological parameter $p_i$ the sensitivities of $S_z$ and $S_v$ are related via
\begin{equation}
\frac{{\partial S_v}/{\partial p_i}}{{\partial S_z}/{\partial p_i}}=\frac{k}{1+z}\,.    
\end{equation}
For our fiducial model we can then find the following sensitivities
\begin{equation}
\frac{\partial S_z}{\partial h}=1+z-E(z)
\end{equation}
\begin{equation}
\frac{\partial S_z}{\partial\Omega_m}=-\frac{h(1+z)^3}{2E(z)}\left[1-(1+z)^{3(w_0+w_a)}\exp{\left[\frac{-3w_az}{1+z}\right]}\right]
\end{equation}
\begin{equation}
\frac{\partial S_z}{\partial w_0}=-\frac{3h(1-\Omega_m)}{2E(z)}(1+z)^{3(1+w_0+w_a)}\ln{(1+z)}\exp{\left[\frac{-3w_az}{1+z}\right]}
\end{equation}
\begin{equation}
\frac{\partial S_z}{\partial w_a}=-\frac{3h(1-\Omega_m)}{2E(z)}(1+z)^{3(1+w_0+w_a)}\left[\ln{(1+z)}-\frac{z}{1+z}\right]\exp{\left[\frac{-3w_az}{1+z}\right]}\,.
\end{equation}
Note that the sign of the ${\partial S_z}/{\partial h}$ term will depend on redshift, while those of the other derivatives are always negative for observationally reasonable values of the model parameters.

Some sensitivity ratios are also illuminating. Consider that of the two dark energy equation of state parameters
\begin{equation}
\frac{{\partial S_z}/{\partial w_a}}{{\partial S_z}/{\partial w_0}}=1-\frac{z}{(1+z)\ln{(1+z)}}\,;
\end{equation}
as one would expect this approaches zero as $z\to 0$ and unity as $z\to \infty$. On the other hand, comparing the sensitivities to the matter density and the present-day dark energy equation of state one finds
\begin{equation}
\frac{{\partial S_z}/{\partial \Omega_m}}{{\partial S_z}/{\partial w_0}}=\frac{1-(1+z)^{3(w_0+w_a)}\exp{\left[\frac{-3w_az}{1+z}\right]}}{3(1-\Omega_m)\ln{(1+z)}(1+z)^{3(w_0+w_a)}\exp{\left[\frac{-3w_az}{1+z}\right]}}\,,
\end{equation}
which tends to $(-w_0)/(1-\Omega_m)$ as $z\to 0$ and to infinity as $z\to \infty$ (though note that in this we are neglecting the radiation density). Considering the specific case of a fiducial $\Lambda$CDM model with $\Omega_m=0.3$, at $z=1$, the ratios are respectively
\begin{equation}
\frac{{\partial S_z}/{\partial w_a}}{{\partial S_z}/{\partial w_0}}=1-\frac{1}{2\ln{2}}\sim0.28
\end{equation}
\begin{equation}
\frac{{\partial S_z}/{\partial \Omega_m}}{{\partial S_z}/{\partial w_0}}\sim\frac{7}{2.1\ln{2}}\sim4.8\,.
\end{equation}

Finally, it is also illuminating to consider the low-redshift limits of the various derivative terms. To obtain them, one uses the fact that as $z\to 0$ we have
\begin{equation}
E(z)=1+\frac{3}{2}\left[1+(1-\Omega_m)w_0\right]z+{\cal O}(z^2)\,.
\end{equation}
For the derivatives with respect to $h$, $\Omega_m$ and $w_0$ we then find, respectively
\begin{equation}
\frac{\partial S_z}{\partial h}\longrightarrow -\frac{1}{2}\left[1+3(1-\Omega_m)w_0\right]z
\end{equation}
\begin{equation}
\frac{\partial S_z}{\partial\Omega_m} \longrightarrow \frac{3}{2}hw_0 z
\end{equation}
\begin{equation}
\frac{\partial S_z}{\partial w_0} \longrightarrow -\frac{3}{2}h(1-\Omega_m)z\,
\end{equation}
while that with respect to $w_a$ (whose value, interestingly, does not affect any of the other three) is of higher order.

\begin{figure}
\begin{center}
\includegraphics[width=\columnwidth,keepaspectratio]{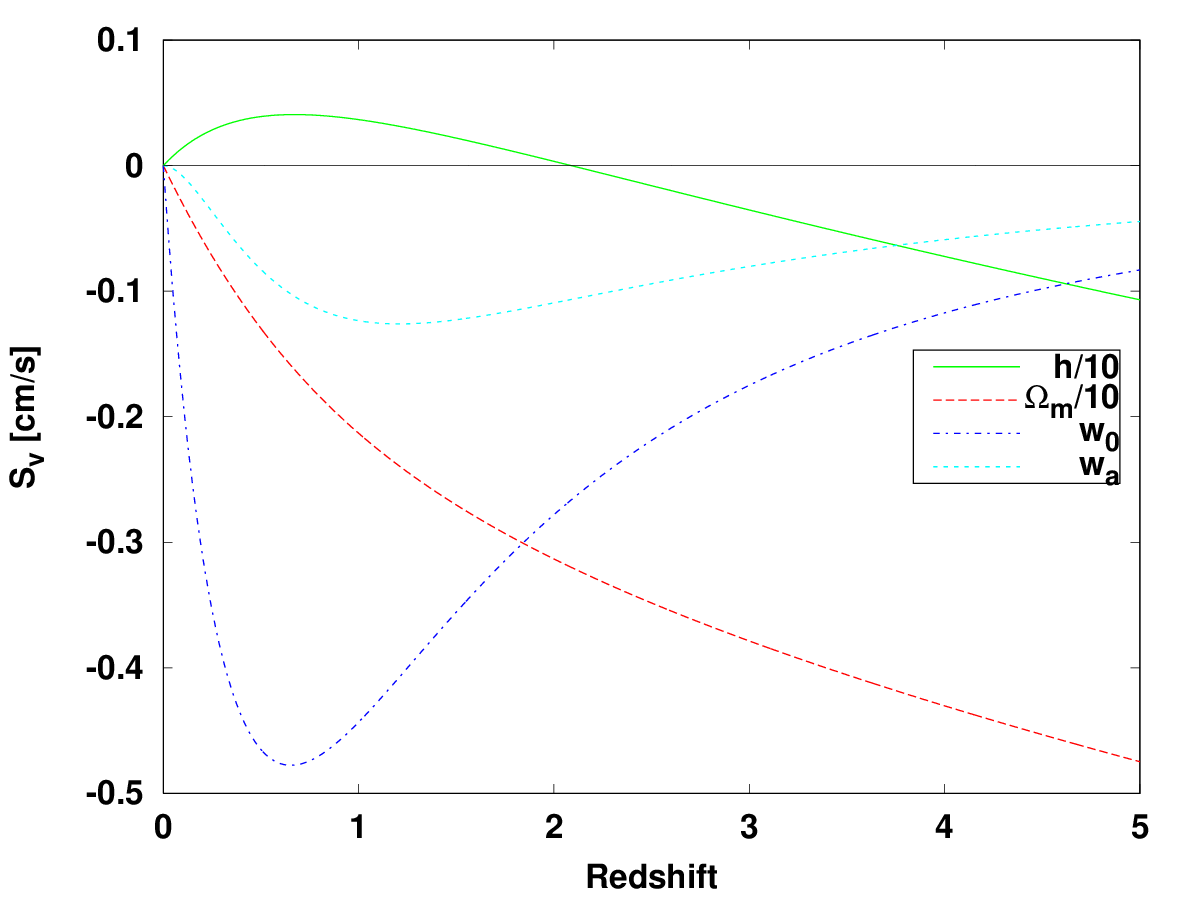}
\includegraphics[width=\columnwidth,keepaspectratio]{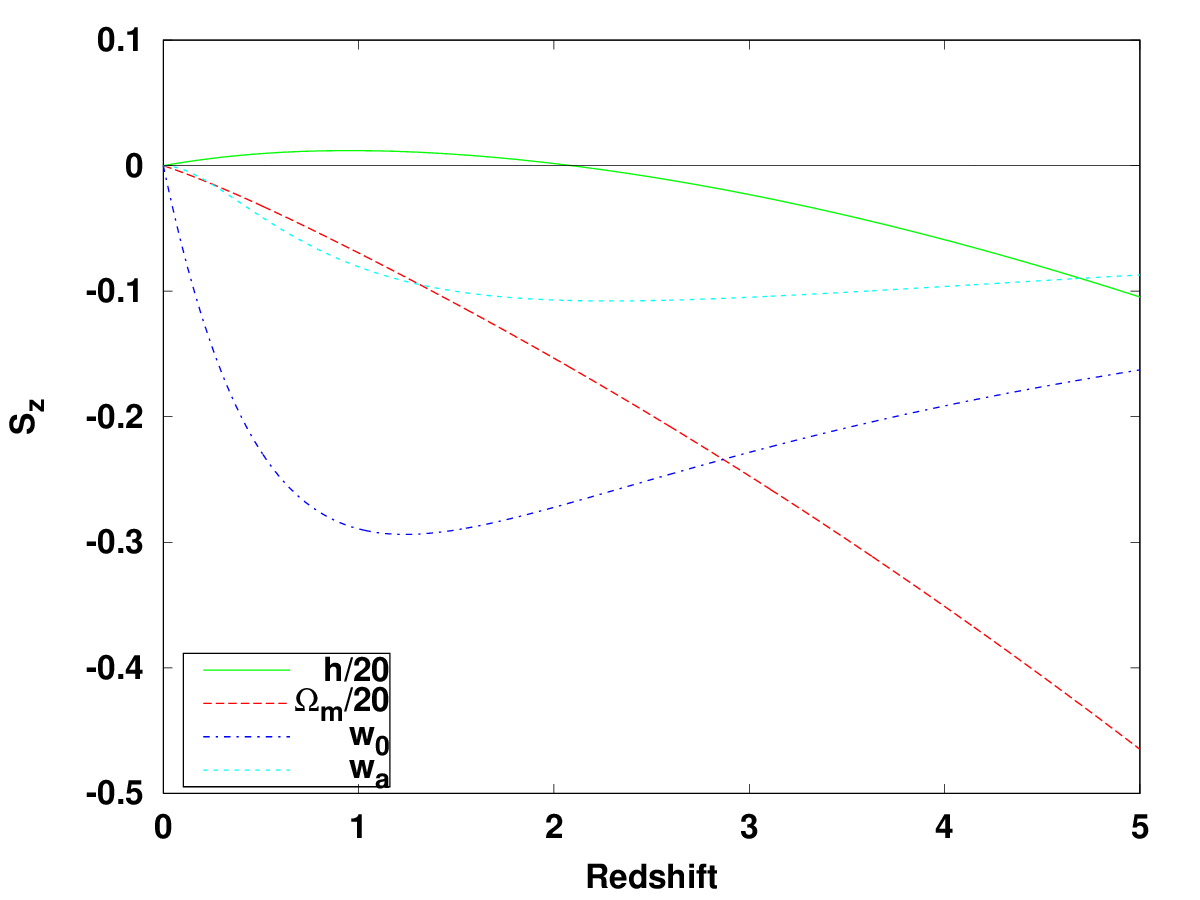}
\end{center}
\caption{Sensitivities of the spectroscopic velocity (top panel) and of the redshift drift signal (bottom panel) to the cosmological parameters in the CPL parametrization, for a flat $\Lambda$CDM fiducial. For plotting convenience the sensitivities to $h$ and $\Omega_m$ have been divided by a factor of 10 in the top panel and by a factor of 20 in the bottom panel. The zero sensitivity line is shown in black. Similar plots can be found in Figure 4 of Ref. \protect\citenum{drift1}.}
\label{fig1}
\end{figure}

Figure \ref{fig1} depicts the redshift dependence of the two observational sensitivities, $\partial S_v/\partial p_i$ and $\partial S_z/\partial p_i$, for a flat $\Lambda$CDM fiducial model with $\Omega_m=0.3$ and $h=0.7$. Note that the two are slightly different, being related by a redshift-dependent factor. Also note the different units in the vertical axes of the two panels: $\partial S_z/\partial p_i$ is dimensionless, while $\partial S_v/\partial p_i$ has units of cm/s.

As expected the sensitivity on the dimensionless Hubble constant $h$ (which is proportional to the redshift drift signal itself) is the only one that changes sign; this and the sensitivity on $\Omega_m$ are also larger than those on the dark energy equation of state parameters $w_0$ and $w_a$. Nevertheless, the most noteworthy point is that the various sensitivity curves have different redshift dependencies. While the matter density sensitivity increases monotonically (in absolute value) with redshift, those of $w_0$ and $w_a$ are maximal at around the onset of acceleration (just below $z=1$, with the value being slightly different for $S_v$ and $S_z$), and the former has a stronger redshift dependence than the latter. This is important because it implies that as long as one is able to do these measurements at sufficiently broad redshift ranges there should not be strong covariances between the parameters \cite{drift1}.

\section{Three experimental strategies}

As previously mentioned, redshift drift measurements are unique in comparing different past light cones in a fully model-independent way, rather than mapping our present-day past light cone. Nevertheless, these measurements can complement traditional probes in constraining specific cosmological models. Indeed they have an important role to play here, since they do probe different regions of parameter space, enabling the breaking of key degeneracies upon combination. One may therefore ask how an experimental strategy aimed at maximizing the cosmological constraining power of the redhift drift (on its own or in combination with traditional probes) compares with a strategy aimed at maximizing the statistical significance of the detection of the drift signal itself.

This is a particularly pressing question since two other approaches to detecting the redshift drift have been recently proposed. The Cosmic Accelerometer \cite{Eikenberry} aims to lower the cost of the experiment, while the Acceleration Programme \cite{Cooke} proposes to measure the differential redshift drift between two non-zero redshifts, rather than measuring the drift with respect to today. In Ref. \citenum{drift2} we again used Fisher Matrix techniques in a comparative study of the cosmological impact of the three approaches. The fiducial model choices were also those described above, to maintain consistency with Ref. \citenum{drift1}. Herewith we introduce these alternative strategies and summarize the results of this analysis.

The case study for redshift measurements by the ELT has been done in Ref. \citenum{Liske}. The measurement relies on absorption features in the Lyman-$\alpha$ forest, complemented by metal absorption lines redwards of it. The main practical result of this study is that the spectroscopic velocity uncertainty is well described by
\begin{equation}\label{elt}
\sigma_v=\sigma_e\left(\frac{2370}{S/N}\right)\sqrt{\frac{30}{N_{QSO}}}\left(\frac{1+z_{QSO}}{5}\right)^{-\lambda}\, cm/s\,,
\end{equation}
where the last exponent is $\lambda=1.7$ up to $z=4$ and $\lambda=0.9$ for higher redshifts. Consistently with this and also with the latest top-level requirements for the ELT-HIRES spectrograph \cite{HIRES}, we assume that each measurement has a signal to noise ratio $S/N=3000$ and uses data from $N_{QSO}=6$ quasars; as baseline values we take $\sigma_e=1.35$ and an experiment time $\Delta t=20$ years, but we also explore different choices for these two parameters. The ELT can measure the redshift drift in the approximate redshift range $2.0\leq z\leq 4.5$. The main bottleneck, in addition to the spectrograph stability which is thought not to be a bottleneck \cite{Milakovic}, is the availability of bright quasars providing the required signal to noise in reasonable amounts of telescope time. The discovery of additional bright quasars will improve the experiment feasibility \citep{Boutsia}, so our previously reported ELT analysis \cite{drift2} might be conservative. 

The first alternative, dubbed the Cosmic Accelerometer \cite{Eikenberry} (henceforth CAC) can be crudely thought of as a low-cost version of the ELT experiment, relying on commercial off the shelf equipment. The proposal is part of the Astro2020 decadal survey, and to our knowledge no detailed feasibility study has been carried out so far. We therefore take at face value the spectroscopic velocity uncertainty given in the proposal white paper
\begin{equation}\label{cac}
\sigma_v=\sigma_c\sqrt{\frac{6}{t_{exp}}}\, cm/s\,,
\end{equation}
with the baseline value of $\sigma_c=1.5$ and that of the experiment time $t_{exp}=6$ years; the authors of the decadal survey proposal state that a further advantage of this approach would be a detection of the redshift drift on a shorter time scale than that of the ELT. In any case, we will again explore the impact of different values of these parameters. The redshift ranges probed are, in principle, similar to those for the ELT. 

The second alternative is the Acceleration Programme \cite{Cooke} (henceforth APR), which uses the ELT as described above but proposes to measure the drift between sources at two different non-zero redshifts along the same line of sight. This is to be compared to the standard approach, where one of these redshifts is always $z=0$. The measured quantity in this case will be a spectroscopic velocity difference, which making use of previously introduced definitions we can write as
\begin{equation}
\Delta v_{ir}=(cH_0\Delta t)\left[\frac{E(z_r)}{1+z_r}-\frac{E(z_i)}{1+z_i}\right]\,,    
\end{equation}
where $z_r$ and $z_i$ are the redshifts of the reference and intervening sources, implying that $z_i<z_r$. The rationale for this proposal is that for some choices of redshifts the absolute value of the detected drift signal (or, more precisely, that of the corresponding spectroscopic velocity) can be significantly larger than in the standard approach. Since the absolute value of the spectroscopic velocity uncertainty of the measurement, $\sigma_v$, is determined by the spectrograph's technical specifications, maximizing the value of the signal to be detected leads to an increased statistical significance of the detection.

\section{Cosmological constraints based experimental trade-offs}

Our main comparison diagnostic is the derived uncertainty in the cosmological parameters being considered, specifically the matter density fraction and the CPL dark energy equation of state parameters. Our fiducial model for the comparison is flat $\Lambda$CDM, with $\Omega_m=0.3$, $H_0=70$ km/s/Mpc, $w_0=-1$ and $w_a=0$.

To compare the ELT and CAC it is convenient to assume a single measurement of the redshift drift. Clearly, even for a flat $\Lambda$CDM fiducial model a single redshift drift measurement can't simultaneously constrain the Hubble constant and the matter density, so one adopts ample uniform priors for both of these. We can then explore the impact of choices of normalization of the spectroscopic velocity uncertainties ($\sigma_e$ and $\sigma_c$), of the experiment time, and of the redshift at which the measurement is made. Our specific metric in this case be the posterior matter density constraint, $\sigma_m$.

\begin{figure}
\begin{center}
\includegraphics[width=\columnwidth,keepaspectratio]{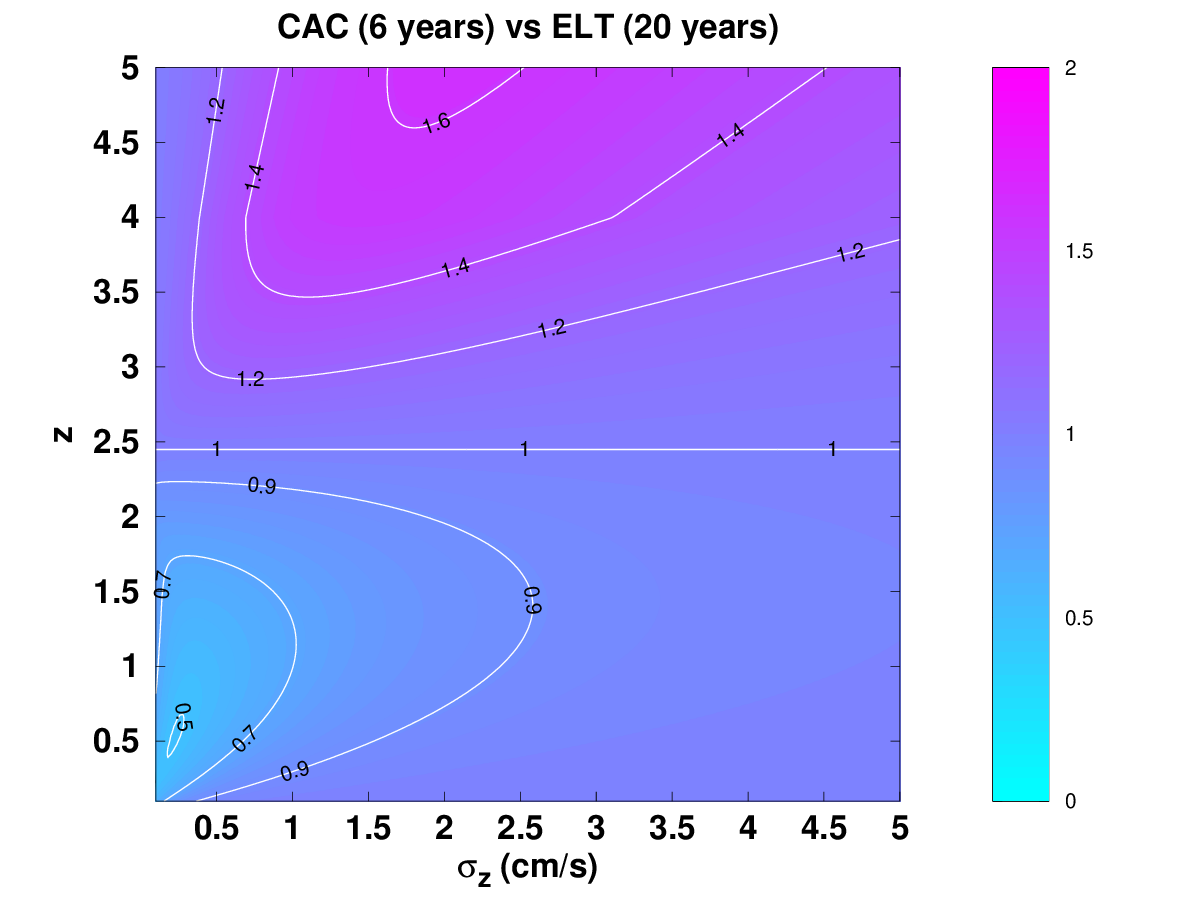}
\includegraphics[width=\columnwidth,keepaspectratio]{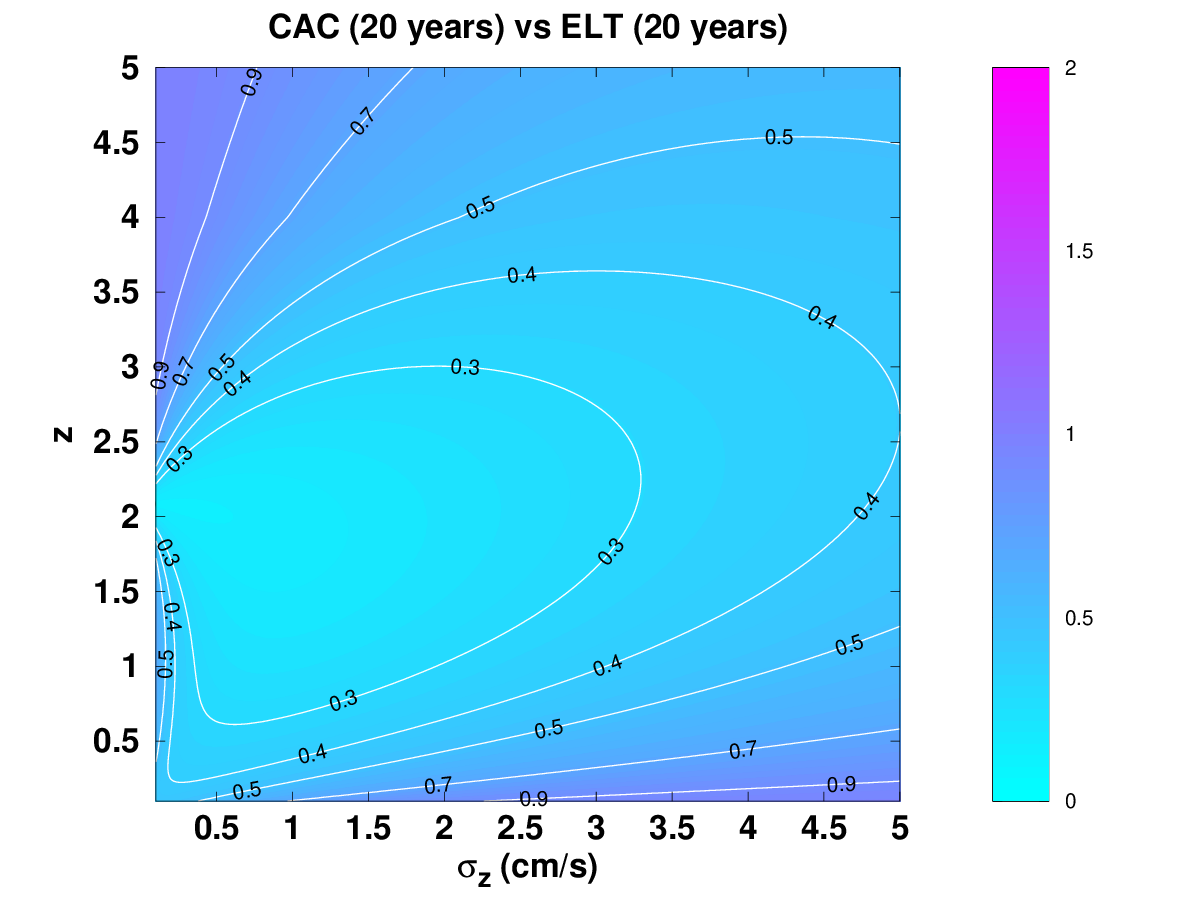}
\end{center}
\caption{Ratio of the constraints on the matter density ($\sigma_m$) from a single redshift drift measurement by the CAC and ELT, as a function of the redshift of the measurement ($z$) and the velocity uncertainty ($\sigma_z$), with a CAC experiment time of 6 or 20 years (top and bottom panels), and 20 years for ELT in both cases. A ratio larger than unity implies that the ELT measurement is more constraining. An alternative depiction of this comparison can be found in Figure 2 of Ref. \citenum{drift2}.}
\label{fig2}
\end{figure}

The crucial point for this analysis, already discussed and illustrated in Fig. \ref{fig1} is that the cosmological parameter sensitivities of the redshift drift are redshift-dependent \cite{drift1}. This sensitivity must therefore be convolved with the spectroscopic velocity uncertainty of each experiment, bearing in mind that the ELT one is redshift dependent (cf. Eq.~\ref{elt}) while the CAC one is not (cf. Eq. \ref{cac}), at least if one takes the CAC proposal at face value.

In a nutshell, the main result of this comparison is that for identical experiment times the CAC does better, though the difference decreases for larger redshifts. A possible advantage of the CAC (in addition to cost) would be a lower experiment time. The top panel of Figure \ref{fig2} compares the constraints on $\sigma_m$ for a 6 year CAC and a 20 year ELT experiment, and one finds that in these circumstances the ELT does clearly better at the redhsifts of interest---up to $62\%$ better. The reason for this is clear. While absolute uncertainties $\sigma_v$ may be smaller for the CAC, the relative ones will be larger due to smaller experiment times, since a key feature of the redshift drift is that the signal grows linearly in time. The bottom panel of the same figure compares both experiments at 20 years; in this case the CAC is more constraining---by up to a factor of six for measurements around redshift $z\sim2$.

We therefore see that an experiment optimized to detect the redshift drift signal is not necessarily optimized to constrain particular cosmological models, or indeed specific model parameters. In the example of Figure \ref{fig2} the constraints on the matter density can be substantially different, but those on the Hubble constant never differ by more than a few percent \cite{drift2}. 

\begin{figure}
\begin{center}
\includegraphics[width=\columnwidth,keepaspectratio]{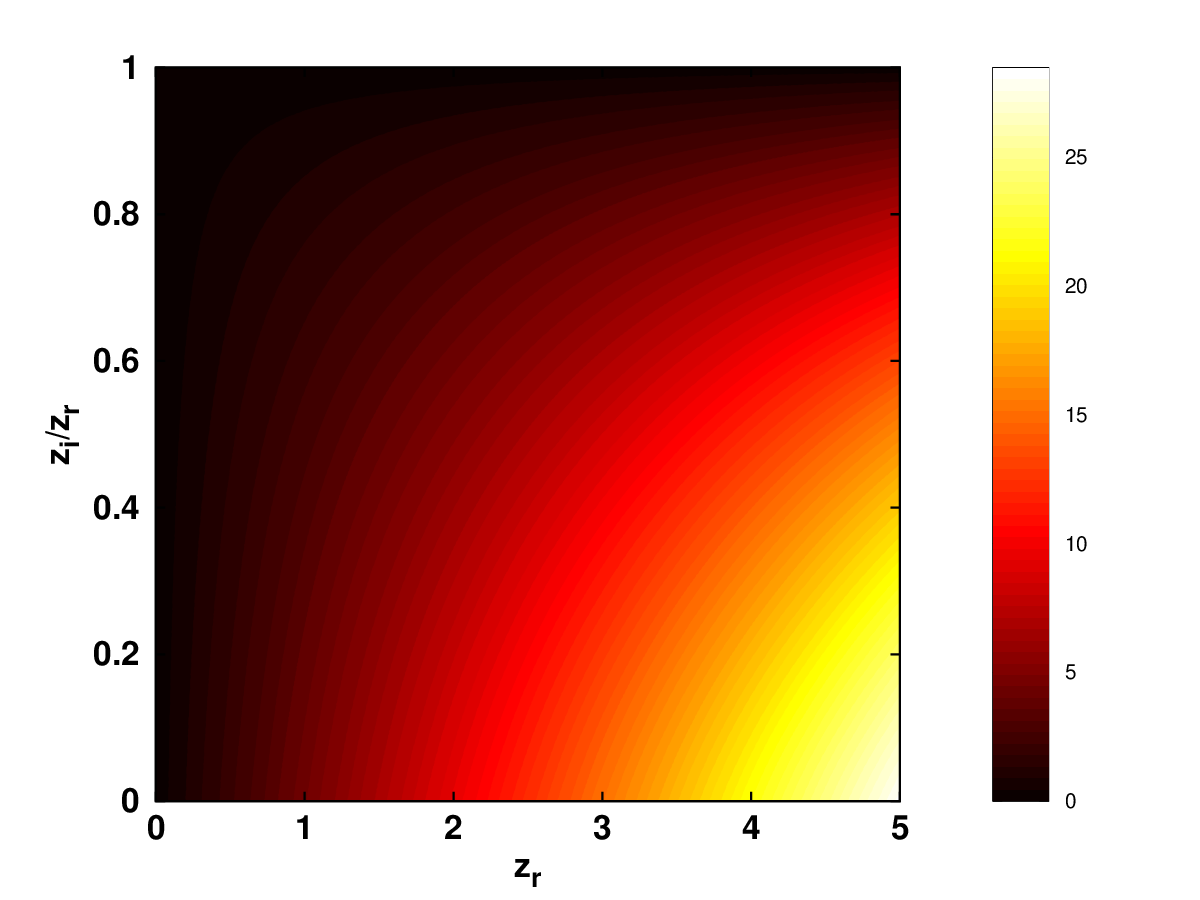}
\includegraphics[width=\columnwidth,keepaspectratio]{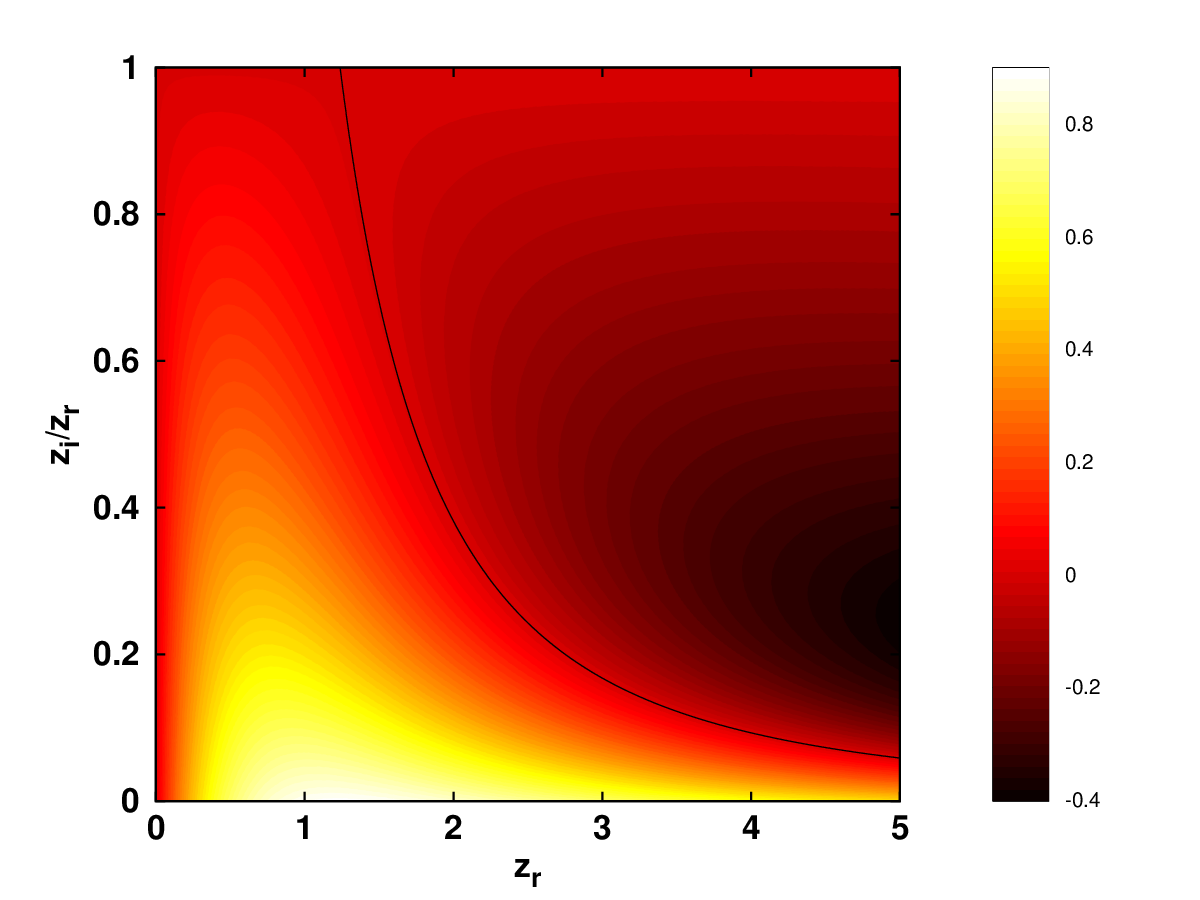}
\end{center}
\caption{The sensitivities of the differential redshift drift to the cosmological parameters $\Omega_m$ (top) and $w_0$ (bottom), as a function of the reference redshift $z_r$ and the ratio of intervening and reference redshifts $z_i/z_r$. The colormaps show the sensitivity in cm/s per year of observation, and the black curves show the locus of zero sensitivity (for $\Omega_m$ there is no such non-trivial locus). An alternative depiction of these sensitivities, also for $h$ and $w_a$, can be found in Figure 3 of Ref. \citenum{drift2}.}
\label{fig3}
\end{figure}

Analogous results ensue when comparing the parameter uncertainties in the canonical redshift drift and the differential redshift drift, for the CPL parameterization. To remain consistent with the work on the canonical redshift drift \cite{drift1}, in Ref. \citenum{drift2} we used a set of five redshift drift measurements at reference redshifts $z_r= \{2.0, 2.5, 3.0, 3.5, 4.5\}$, and a 20 years experiment. As for the choice of intervening redshifts one may consider two representative choices one with all $z_1 =0.67$ (this being the redshift with maximum positive spectroscopic velocity for our choice of fiducial model) and another numerically determined with the goal of maximizing the Figure of Merit for the dark energy equation of state parameters, $w_0$ and $w_a$, defined as the inverse of the area of the one-sigma confidence contour in the two-dimensional parameter space.

The results, further detailed in Ref. \citenum{drift2}, confirm that amplifying the signal for the redshift drift does not necessarily improve the uncertainties of any parameters. However, the numerically determined set of redshifts shows that in ideal circumstances the uncertainties for $w_0$ and $w_a$ can be improved by one to two orders of magnitude with respect to the canonical redshift drift, while the uncertainties for $h$ and $\Omega_m$ were also improved. Nevertheless this advantage is diluted when priors are added, due to relatively high non-diagonal terms in the Fisher matrices. (In particular, we see that constraints on the dark energy equation of state parameters are dominated by the priors, making the ratios of the corresponding uncertainties become unity.) It will be important to validate these results though a MCMC analysis.

These results can be understood by calculating the sensitivity coefficients of the differential redshift drift to our four cosmological parameters, $\partial (\Delta v_{ir})/\partial p$, where $p$ is one of $(h,\Omega_m,w_0,w_a)$, These are analogous to the ones shown in Fig. \ref{fig1} for the standard redshift drift, except that they now depend on both the reference and intervening redshifts. As expected, one finds that the sensitivity to $\Omega_m$ is much larger than the others. Crucially, the sensitivity to $\Omega_m$ is maximized for $z_i=0$ and maximal $z_r$: in other words, if the goal is to constrain $\Omega_m$, the canonical redshift drift measurement is always the optimal strategy. On the other hand, for the other model parameters the differential redshift drift can provide tighter constraints.

\section{Synergy with classical cosmological probes}

Throughout this analysis, it has become clear that optimal constraints on the background quantities dynamically involved in the redshift-drift require wide redshift intervals to take advantage of the individual parameter sensitivities over the expansion history. This justifies, inter alia, the complementarity of redshift drift experiments, whose main candidates are embodied by the phase-2 SKA for low-redshift spiral galaxies and the ELT for distant quasars. Indeed, as previously pointed out, not only do these experiments complement each other, but there are also hints that they present promising synergies with traditional cosmological probes, potentially allowing to achieve more stringent joint constraints over cosmological parameters of the concordance model and its extensions \cite{drift1}. Specifically, the fundamentally different nature of the redshift-drift experiment with respect to CMB, SNe, galaxy clustering, or weak lensing experiments, allows for novel correlations within the relevant parameter space, witnessed by degeneracy breaking. As a consistency check and to gain further insights on the degeneracies of the dark energy parameters, it is of great interest to complement current Fisher forecasts with a more thorough MCMC analysis for accurate sampling of the distributions.

Similarly to the previous section, we account for a possible dynamical behavior of dark energy with two degrees of freedom, $w_0$ and $w_a$, through the CPL parametrization. The mock dataset for the ELT is generated according to the prescriptions of the third section for 30 QSOs populating 5 redshift bins between $z=2$ and $z=4.5$. As for the total uncertainty over the simulated data, Eq. \eqref{elt} is used with a somewhat more optimistic value of unity for $\sigma_e$, reflecting the fact that redshifts are measured only at two distinct times (the underlying motivation is to compare the full potential of the redshift-drift against tight CMB and SNe constraints). Moreover, random noise is generated from a Gaussian distribution around the theoretical redshift-drift value, with a standard deviation of $\sigma_v$ to mimic genuine measurements from Lyman-$\alpha$ forests. As for the phase-2 SKA galaxy sample, 5 bins are generated between redshifts 0.2 and 1.0 with redshift-dependent relative uncertainties of 10\% following recommendations of previous estimations \cite{Second}, to which is added the same artificial Gaussian noise over the theoretical signal.

We implemented the theoretical redshift-drift signal to the dynamical landscape at the background level within the Boltzmann solver CLASS \cite{CLASS}. The theoretical drift is constrained to follow Planck 2018's cosmology \cite{Planck18} not to bias the relative distributions and to ease the comparison. The true posterior is then sampled with the Bayesian sampler MontePython \cite{MontePython} in which we integrated the simulated datasets and the scripts to generate the associated Gaussian likelihoods. The priors over all relevant cosmological parameters are consistent with the ones used in the Planck analysis.

\begin{figure}
    \centering
    \includegraphics[width=\columnwidth,keepaspectratio]{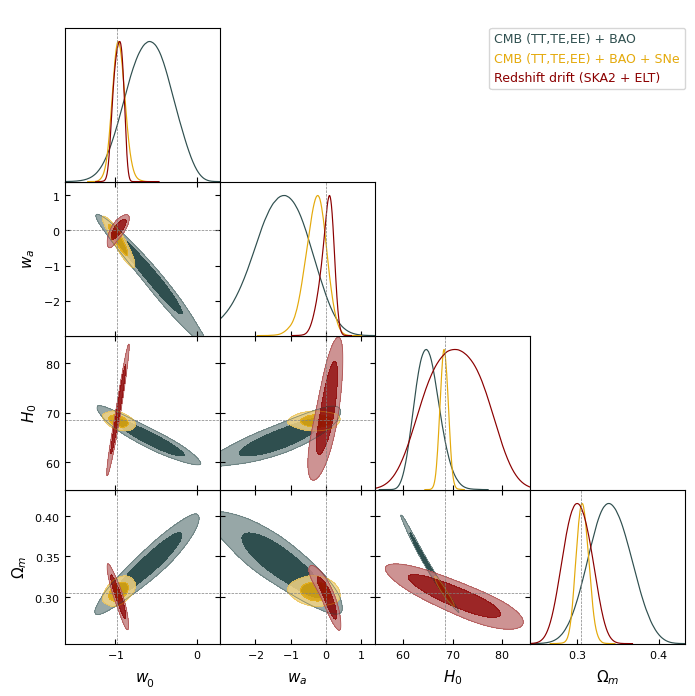}
    \caption{Marginalized posterior distributions in the space $\{\Omega_m, H_0, w_0, w_a\}$ for the joint contribution of redshift drift experiments (red), CMB experiment with BAO measurement (teal) and CMB + BAO + supernovae (yellow). The data considered for the redshift drift experiments are the mock SKA2 and ELT samples whereas classical probes (CMB, BAO, SNe) are respectively taken from Planck, BOSS DR12, and the Pantheon supernovae samples.}
    \label{fig4}
\end{figure}

The most noteworthy results from the forecast illustrated in Fig.\ref{fig4} are first that there is no significant deviation from Gaussianity within the distribution, which further legitimates previous Fisher analyses. More importantly, one can note the remarkable degeneracy breaking of the redshift drift probability density function on most of the parameters with respect to the other probes. In particular, the clear degeneracy within the $w_0$--$w_a$ space tends to suggest that the joint analysis of classical probes with future redshift drift datasets could further improve the constraints on the first-order term of the CPL parametrization and therefore work toward a better characterization of the possible dynamical behavior of dark energy. Here, the contribution of the low redshift SKA2 sample is important to this endeavor as it probes the dark energy-dominated era and allows to reveal a positive correlation between the two parameters whereas $w_0$ and $w_a$ are anticorrelated for the other current cosmological observables. Nevertheless, one should keep in mind that such a combination of datasets requires particular attention to the possible correlations between probes and individual systematics.

That being said, in the light of these results, we should expect a consistent gain in the constraints on $\Omega_m$, $w_0$, and $w_a$ with this type of configuration for SKA and ELT from the joint analysis of the redshift drift measurements with current cosmological probes. In the near future, a better instrumental characterization of these facilities is likely to refine the present conservative forecasts, and the CAC program, as well as intensity mapping experiments, will provide more data samples and homogenize the covered redshift range. Furthermore, by the time of the first observations of the redshift drift, next-generation galaxy surveys (Euclid, DESI), supernovae surveys (Vera C. Rubin Observatory) but also CMB experiments (Simons observatory, CMB-S4) will improve individual constraints, although in absence of new observable to present different correlations, these experiments will still highly benefit from the redshift drift degeneracy breaking.

\section{Conclusions and outlook}

The era of redshift drift measurements (rather than upper limits) is coming. These experiments need to be optimized for a specific purpose, which might be the detection of the drift signal itself, or constraining specific model parameters, such as the matter density or the energy equation of state, or providing a definitive consistency test for the model-dependent analysis of traditional cosmological datasets. Our analysis shows that optimized observational strategies for each of these are significantly different.

We note that an assumption of our analysis is that the measurement redshifts are free parameters, thus ignoring instrumental and observational limitations. In practice one can only do the measurements with the known bright quasars and their absorption systems. The number of astrophysical targets for these measurements remains relatively small despite recent progress \cite{Liske,Boutsia}. The Phase B of construction of the ELT-HIRES spectrograph, which will likely do the first redshift drift measurements, is about to start, and it will include a detailed feasibility study that will further explore the issues raised herein.

\section*{Acknowledgments}

This work was financed by FEDER---Fundo Europeu de Desenvolvimento Regional funds through the COMPETE 2020---Operational Programme for Competitiveness and Internationalisation (POCI), and by Portuguese funds through FCT - Funda\c c\~ao para a Ci\^encia e a Tecnologia in the framework of the project POCI-01-0145-FEDER-028987 and PTDC/FIS-AST/28987/2017. This work was partially enabled by funding from the UCL Cosmoparticle Initiative.

\eject

\bibliographystyle{ws-procs961x669}
\bibliography{martinsdrift}

\begin{thebibliography}{10}

\bibitem{drift1}
C.~S. Alves, A.~C.~O. Leite, C.~J. A.~P. Martins, J.~G.~B. Matos and T.~A.
  Silva, {Forecasts of redshift drift constraints on cosmological parameters},
  {\em Mon. Not. Roy. Astron. Soc.} {\bf 488}, 3607  (2019).

\bibitem{drift2}
J.~Esteves, C.~J. A.~P. Martins, B.~G. Pereira and C.~S. Alves, {Cosmological
  impact of redshift drift measurements}, {\em Monthly Notices of the Royal
  Astronomical Society: Letters} {\bf 508}, L53 (08 2021).

\bibitem{Sandage}
A.~{Sandage}, {The Change of Redshift and Apparent Luminosity of Galaxies due
  to the Deceleration of Selected Expanding Universes.}, {\em Ap. J.} {\bf
  136}, p. 319  (1962).

\bibitem{Mcvittie}
G.~C. {McVittie}, {Appendix to The Change of Redshift and Apparent Luminosity
  of Galaxies due to the Deceleration of Selected Expanding Universes.}, {\em
  Ap. J.} {\bf 136}, p. 334  (1962).

\bibitem{Darling}
J.~Darling, {Toward a Direct Measurement of the Cosmic Acceleration}, {\em
  Astrophys. J.} {\bf 761}, p. L26  (2012).

\bibitem{HIRES}
J.~Liske {\em et~al.}, {\em {Top Level Requirements For ELT-HIRES}}, tech.
  rep., Document ESO 204697 Version 1  (2014).

\bibitem{Liske}
J.~Liske {\em et~al.}, {Cosmic dynamics in the era of Extremely Large
  Telescopes}, {\em Mon. Not. Roy. Astron. Soc.} {\bf 386}, 1192  (2008).

\bibitem{Klockner}
H.-R. Klockner, D.~Obreschkow, C.~Martins, A.~Raccanelli, D.~Champion, A.~L.
  Roy, A.~Lobanov, J.~Wagner and R.~Keller, {Real time cosmology - A direct
  measure of the expansion rate of the Universe with the SKA}, {\em PoS} {\bf
  AASKA14}, p. 027  (2015).

\bibitem{Chime}
H.-R. Yu, T.-J. Zhang and U.-L. Pen, {Method for Direct Measurement of Cosmic
  Acceleration by 21-cm Absorption Systems}, {\em Phys. Rev. Lett.} {\bf 113},
  p. 041303  (2014).

\bibitem{HIRAX}
L.~B. Newburgh {\em et~al.}, {HIRAX: A Probe of Dark Energy and Radio
  Transients}, {\em Proc. SPIE Int. Soc. Opt. Eng.} {\bf 9906}, p. 99065X
  (2016).

\bibitem{Quercellini}
C.~Quercellini, L.~Amendola, A.~Balbi, P.~Cabella and M.~Quartin, {Real-time
  Cosmology}, {\em Phys. Rept.} {\bf 521}, 95  (2012).

\bibitem{Second}
C.~J. A.~P. Martins, M.~Martinelli, E.~Calabrese and M.~P. L.~P. Ramos,
  {Real-time cosmography with redshift derivatives}, {\em Phys. Rev.} {\bf
  D94}, p. 043001  (2016).

\bibitem{FMA1}
A.~Albrecht {\em et~al.}, {Report of the Dark Energy Task Force, arXiv:
  astro-ph/0609591},  (2006).

\bibitem{FMA2}
A.~Albrecht {\em et~al.}, {Findings of the Joint Dark Energy Mission Figure of
  Merit Science Working Group, arXiv: astro-ph/0901.0721},  (2009).

\bibitem{CPL1}
M.~Chevallier and D.~Polarski, {Accelerating universes with scaling dark
  matter}, {\em Int. J. Mod. Phys.} {\bf D10}, 213  (2001).

\bibitem{CPL2}
E.~V. Linder, {Exploring the expansion history of the universe}, {\em Phys.
  Rev. Lett.} {\bf 90}, p. 091301  (2003).

\bibitem{Eikenberry}
S.~S. Eikenberry {\em et~al.}, {Astro2020 Project White Paper: The Cosmic
  Accelerometer} (7, 2019).

\bibitem{Cooke}
R.~Cooke, {The ACCELERATION programme: I. Cosmology with the redshift drift},
  {\em Mon. Not. Roy. Astron. Soc.} {\bf 492}, 2044  (2020).

\bibitem{Milakovic}
D.~Milakovi\'c, L.~Pasquini, J.~K. Webb and G.~Lo~Curto, {Precision and
  consistency of astrocombs}, {\em Mon. Not. Roy. Astron. Soc.} {\bf 493}, 3997
   (2020).

\bibitem{Boutsia}
K.~Boutsia, A.~Grazian, G.~Calderone, S.~Cristiani, G.~Cupani, F.~Guarneri,
  F.~Fontanot, R.~Amorin, V.~D’Odorico, E.~Giallongo and et~al., The
  spectroscopic follow-up of the qubrics bright quasar survey, {\em The
  Astrophysical Journal Supplement Series} {\bf 250}, p.~26 (Sep 2020).

\bibitem{CLASS}
J.~Lesgourgues, The cosmic linear anisotropy solving system (class) i: Overview
   (2011).

\bibitem{Planck18}
N.~Aghanim, Y.~Akrami, M.~Ashdown, J.~Aumont, C.~Baccigalupi, M.~Ballardini,
  A.~J. Banday, R.~B. Barreiro, N.~Bartolo and et~al., Planck 2018 results,
  {\em Astronomy \& Astrophysics} {\bf 641}, p.~A6 (Sep 2020).

\bibitem{MontePython}
T.~Brinckmann and J.~Lesgourgues, Montepython 3: boosted mcmc sampler and other
  features  (2018).

\end{thebibliography}

\end{document}